# Precision Predictions

Ulf-G. Meißner [1,2,3,†]

[1]*Helmholtz-Institut für Strahlen- und Kernphysik and Bethe Center for Theoretical Physics, Universität Bonn, D-53115 Bonn, Germany*
[2]*Forschungszentrum Jülich, Institute for Advanced Simulation, Institut für Kernphysik and Jülich Center for Hadron Physics, D-52425 Jülich, Germany*
[3] *Tbilisi State University, 0186 Tbilisi, Georgia*

[†]*Electronic address:* meissner@hiskp.uni-bonn.de

Precision predictions combined with precise measurements are a major tool in sharpening our understanding of the fundamental laws underlying microscopic as well as macroscopic systems. Here, I present a few remarkable examples covering the fields of nuclear, particle and astrophysics.

## 1. PROLOGUE

This manuscript grew out of a talk at the first Joint ECFA-NuPECC-ApPEC Seminar at Paris in October 2019 (JENAS-2019) that brought together physicists from particle, hadronic and nuclear physics as well as from astrophysics and cosmology. I was asked to summarize the role of ``precision predictions'' at this meeting. Clearly, given the time constraints, this could only cover a very small fraction of all the intriguing results in the different fields and the choice of topics therefore had to be entirely subjective.

## 2. INTRODUCTION

First, I should define what is meant by a precision prediction: *A prediction is considered precise, if it has a small (relative) theoretical uncertainty.* This, however, does not imply that it agrees with experiment. Also, the mentioned small uncertainty can be best quantified if we have an underlying counting rule based on some small parameter. Needless to say that a prediction without uncertainty makes little sense. Finally, in what follows I will mostly consider the interplay of precision predictions with the corresponding precise experiments.

To set the stage, let me consider two by now classical examples. The first one concerns the masses of the top quark and the Higgs boson, that where already known within certain ranges before the direct measurements. The underlying idea is that virtual heavy particles can leaves traces in processes involving lighter ones, such as the top quark in loops in electron-positron collisions at LEP producing e.g. a $b\bar{b}$ pair that further hadronizes into jets, see Fig.1.

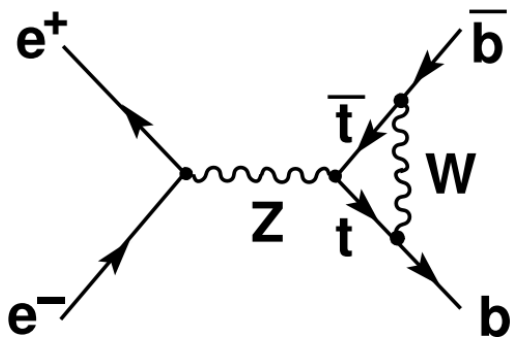

FIG. 1: A typical one-loop radiative corrections in $e^+e^- \to b\bar{b}$ at LEP with a virtual top quark excitation.

Combining such precision measurements with extremely precise higher-loop electroweak calculations, the top quark mass was known to be in the range between 150 and 200 GeV, see e.g. [1], completely consistent with the direct measurement of 175 GeV at the Tevatron in 1995. Similarly, a window for the Higgs boson mass was set by radiative corrections between 114 and 156 GeV, again consistent with the direct measurement of 125 GeV at CERN in 2012. The second well-known example is the so-called Hulse-Taylor pulsar, which has led to fine tests of general relativity (GR). In fact, GR predicts the slowing down of the pulsar period $P_b$ due to the radiation of gravitational waves. In their famous paper from 1979, Taylor, Fowler and McCullogh state that measurements of relativistic effects in the orbit of this binary pulsar lead to a quantitative confirmation of the existence of gravitational radiation at the level predicted by GR [2]. Over four decades, this system has become a true precision test of GR, clearly giving evidence to the existence of gravitational waves as predicted by GR [3]. Gravitational waves where finally detected directly by LIGO/VIRGO in 2015. In what follows, I discuss a few selected recent results.

## 3. FROM SCHWINGER'S TOMBSTONE TO ULTRAHIGH PRECISION

Dirac made the famous prediction that the Landé factor of an electron is g=2, which was challenged by experiments in the late 1940ties. Schwinger, one of the fathers of QED, did the first calculation of the anomalous magnetic moment of the electron, $a_e = (g_e-2)/2 = \alpha/(2\pi) = 1.1 \cdot 10^{-3}$, with $\alpha$ the fine-structure constant, $\alpha = 1/137.03599\ldots$ [4]. This was the dawn of the precision era and this textbook result is engraved on Schwinger's tombstone in Pasadena. In fact, the anomalous magnetic moment is the most precise prediction of the tremendously successful Standard Model (SM) of the strong, electromagnetic and weak interactions, $a_e = a_e(QED) + a_e(weak) + a_e(strong)$. The QED part splits into various pieces depending on the lepton mass ratios, $a_e(QED) = A_1 + A_2(m_e/m_\mu) + A_2(m_e/m_\tau) + A_3(m_e/m_\mu, m_e/m_\tau)$, in terms of the electron, the muon and the tau mass. To achieve sub-ppb precision as in experiment, we must know $A_1$ to tenth order, as $(\alpha/\pi)^5 = 0.07 \cdot 10^{-12}$. The completed and correct calculation of the 12762 tenth order diagrams was reported by Kinoshita and collaborators in 2018 [5]. The SM prediction reads $a_e(th'y) = 1159652182.037(11)(12)(229) \cdot 10^{-12}$, where the first error is due to QED, the second one stems from QCD and the last one from the uncertainty in $\alpha$, measured from the Rydberg levels in atomic Cs [6]. The weak contribution is too small to feature here. This result agrees remarkably well with the so far most precise measurement, $a_e(exp) = 1159652181.73(28) \cdot 10^{-12}$ [7].

One notices a small tension, but before speculating about a possible beyond the SM contribution, the planned improved measurements of the Rydberg constant and of the anomalous magnetic moment of the electron should be performed, see e.g. [8].

The effects of heavy mass particles are enhanced by $(m_\mu/m_e)^2 \sim 43000$ in the muon (g-2). Here, there is a tension between the most precise experiment and the theory, see e.g. [9], but to really draw firm conclusions, one is eagerly awaiting on the one hand the result of the new Fermilab measurement and on the other hand an improved calculation of the theoretical uncertainty for the hadronic light-by-light scattering contribution based on dispersion theory [10].

## 4. PRECISION SIGMA-TERM PHYSICS

Massless classical QCD is invariant under scale transformations (dilatations), $r \to \lambda r$, with $\lambda$ a real number. This scale invariance is broken by quantization, the well-known dimensional transmutation leads to the scale $\Lambda_{QCD}$ = 250 MeV, that can e.g. be inferred from the running of the strong coupling constant as well as the non-vanishing of the trace of the QCD energy-momentum tensor, $\Theta^\mu_\mu$. This so-called *trace anomaly* leads to the generation of hadron masses. For the proton |p>, this reads (neglecting a small anomalous dimension term) [11]

$$m_p = <p|\Theta^\mu_\mu|p>$$
$$= <p|(\beta_{QCD}/g)G^a_{\mu\nu}G^{\mu\nu}_a + m_u \bar{u}u + m_d \bar{d}d + m_s \bar{s}s|p>, \quad (1)$$

with $\beta_{QCD}$ the QCD $\beta$-function, g the strong coupling constant, $G^{\mu\nu}_a$ the gluon field strength tensor and $m_f$ the mass of the quark with flavour f, f=u,d,s. The first term in Eq.(1) is pure gluon field energy and the last three terms give the contribution from the Higgs boson to the proton mass. The term proportional to the light up and down (strange) quark masses is called the pion-nucleon (strangeness) sigma-term, $\sigma_{\pi N}$ and $\sigma_s$, respectively. These sigma-terms play a much larger role than just giving a part of the proton mass, they parameterize the scalar couplings of the nucleon, that are of utmost importance for direct dark matter detection as well as muon to electron conversion in nuclei. The pion-nucleon sigma-term also features in the leading density-dependence of the scalar quark condensate and in CP-violating $\pi N$ couplings that contribute to electric dipole moments of the nucleon and light nuclei.

The $\pi N$ sigma-term can most precisely be determined from a Roy-Steiner (RS) analysis of pion-nucleon scattering, using the precision pionic atom data from PSI that allow for an accurate extraction of the $\pi N$ S-wave scattering lengths [12]. The RS analysis is in fact the first ever dispersive analysis of $\pi N$ scattering with error bars, it leads to a high-precision determination of $\sigma_{\pi N}$ =59.1(3.5) MeV [13], which is quite an achievement in hadron physics. The strangeness sigma-term is less well determined, combining lattice QCD and chiral perturbation theory results leads to $\sigma_s$ = 30(30) MeV. Consequently, only about 100 MeV of the nucleon mass are due to the Higgs. Stated differently, in a world with massless quarks, the proton would still weigh in with about 840 MeV, quite different from the pion, that would be massless in such a world due to its Goldstone boson nature. This is a central result of QCD! It should be noted, however, that present lattice QCD determinations of the pion-nucleon sigma-term are inconsistent with our knowledge of the S-wave pion-nucleon scattering lengths, see [14]. This is definitely a challenge to the lattice QCD community.

## 5. PRECISION SHAPIRO DELAY PHYSICS

In his seminal paper in 1915, Einstein proposed three tests of GR, namely the perihelion motion of Mercury, the bending of light by massive bodies and gravitational waves. While the first was already observed earlier, light bending was seen by Eddington in 1919 and gravitational waves as predicted by GR were detected in 2015. In 1964, Shapiro had proposed a fourth test of GR, namely the time delay in a signal due to the reduction of the speed of light in curved space-time [15]. More precisely, light moves on geodesics and these are modified in curved space-time, leading to a delay in the arrival time of a signal. In the standard approximation for a binary system, where the signal is sent from the body A, the Shapiro delay is given in terms of two parameters (first post-Newtonian approximation), namely the Shapiro range, $r_{sh}$ = $Gm_B/c^3$, with G Newton's constant, $m_B$ the mass of the companion and c the speed of light, and the so-called Shapiro shape, $s_{sh}$= sin i, with i the inclination of the orbit. To this order in the expansion of $(v/c)^2$, with v the velocity of the companion, the predictions of GR agree very well with the data from the double pulsar PSR J0737-3039A and B [16]. An update of this work is depicted in Fig.2, where new measurements of the Shapiro delay in this system are shown. The delay is largest at superior conjunction (orbital phase of 90 degrees), when the emitting pulsar is located behind its companion as seen from Earth. The solid curve shows the expectation from GR. The remaining small deviations resulting from higher order effects have been detected and will be discussed by an upcoming work in [17]. Higher order propagation delays in the $(v/c)^2$ expansion for a binary system are: 1) retardation [18], 2) light bending [19] and 3) the pulsar rotation [20]. This fine prediction of GR is in remarkable agreement with the most recent data [17]. In fact, the theoretical uncertainty in the prediction of the Shapiro delay is about 0.02%, which is quite amazing.

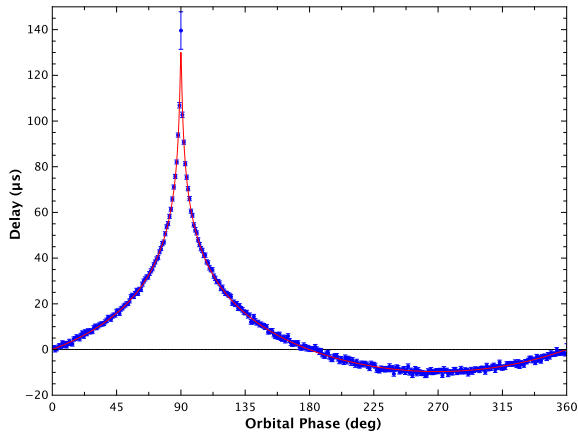

FIG. 2: Shapiro time delay measured for the double pulsar system (PSR J0737-3039A/B) as a function of the orbital phase. Figure courtesy of Michael Kramer.

## 6. PRECISION MEETS ANTHROPICS

Nuclear lattice effective field theory is a relatively new tool to perform *ab initio* nuclear structure and reaction calculations at the sub-percent level, see e.g. the recent monograph [21]. In an earlier article in this journal [22], I had already discussed how this framework can be used to investigate the closeness of the so-called Hoyle state in $^{12}$C to the triple-alpha threshold as a function of the fundamental parameters of the SM. This energy difference plays an eminent role in the discussion of the so-called anthropic principle. I just would like to mention here that there has been a recent update on the calculation of the parameter-dependence of this quantity, partly triggered by new stellar simulations investigating the dependence of carbon and oxygen production in stars on the aforementioned energy difference as well as on the star's metallicity [23]. Also, the description of the quark mass dependence of the NN S-wave scattering lengths, that feature prominently in this calculation, has been improved, but still lattice QCD simulations closer to the physical point are needed to reduce the ensuing uncertainties (even worse, there is some sizeable tension between the existing lattice QCD determinations of the NN S-wave scattering lengths). The most interesting finding of this new work is that the scenario of no fine-tuning in the light quark masses can now be excluded. The interested reader is referred to [24] for a more detailed discussion and further references.

## 7. SUMMARY AND CONCLUSIONS

Let me briefly summarize the main lessons learned here:

1) Precision predictions rest on scale separations, therefore effective field theories are the best tool to make precision predictions (or other methods that can either deal with perturbative or non-perturbative physics in a systematic way).
2) Precision predictions, or, more generally, precision physics, are (is) of ever growing importance.
3) Precision physics is arguably our best take on discovering physics beyond the Standard Model.
4) Consequently, we need to sharpen the predictions where the SM gives only a tiny contribution, such as the electric dipole moments of nucleons and light nuclei.

I hope that with this short essay I could convey the fascination related to precision physics. It sometimes might take a long time that such a prediction can be confronted with an equally precise experiment, but this should not stop us from investing more time and effort into this rewarding field, independent of the area of research one is working in.

### Acknowledgements

I would like to thank all my collaborators sharing their insight into the topics discussed here. I am very grateful to Michael Kramer and Norbert Wex for teaching me the beautiful physics related to the Shapiro delay and double pulsar systems. Computational resources were provided by the Jülich Supercomputer Centre and the RWTH Aachen. Work supported in part by the DFG, the Chinese Academy of Sciences and the VolkswagenStiftung.

### REFERENCES

[1] C. Quigg, Ann. Rev. Nucl. Part. Sci. 59 (2009) 505.
[2] J.H. Taylor et al., Nature 277 (1979) 437.
[3] J.M. Weisberg and Y. Huang, Astrophys.J. 829 (2016) 55.
[4] J. Schwinger, Phys. Rev. 73 (1948) 416.
[5] T. Aoyama et al., Phys. Rev. D97 (2018) 036001.
[6] R.H. Parker et al., Science 360 (2018) 191.
[7] D. Hanneke et al., Phys. Rev. Lett. 100 (2008) 120801.
[8] G. Gabrielse et al., Atoms 7 (2019) 45.
[9] A. Keshavarzi et al., Phys. Rev. D97 (2018) 114025.
[10] G. Colangelo et al., JHEP 1509 (2015) 074.
[11] J.C. Collins et al., Phys. Rev. D16 (1977) 438.
[12] V. Baru et al., Nucl. Phys. A872 (2011) 69.
[13] M. Hoferichter et al., Phys. Rev. Lett. 115 (2015) 092301.
[14] M. Hoferichter et al., Phys. Lett. B760 (2016) 74.
[15] I.I. Shapiro, Phys. Rev. Lett. 13 (1964) 789.
[16] M. Kramer et al., Science 314 (2006) 97.

[17] M. Kramer et al., to be submitted.
[18] R. Blandford and S.A. Teukolsky, Astrophys. J. 205 (1976) 580.
[19] P. Schneider et al., *Gravitational Lenses*, Springer Verlag, Heidelberg (1992).
[20] O.V. Doroshenko and S.M. Kopeikin, Mon. Not. R. Astron. Soc. 274 (1995) 1029.
[21] T. Lähde and U.-G. Meißner., Lect. Notes Phys. 957 (2019) 1.
[22] U.-G. Meißner, Nucl. Phys. News 24 (2014) 11.
[23] L. Huang et al., Astropart. Phys. 105 (2019) 13.
[24] T. Lähde et al., arXiv:1906.00607, *acc. for publication* in Eur. Phys. J. A (2020).